\documentclass[aps,%
a4paper,%
12pt,%
final,%
notitlepage,%
oneside,%
onecolumn,%
nobibnotes,%
nofootinbib,%
superscriptaddress,%
noshowpacs,%
centertags]{revtex4}

\begin{document}

\title{Study of Proton and Deuteron Pickup Reactions
  \(\isotope[2]{H}(\isotope[10]{Be},\isotope[3]{He})\isotope[9]{Li}\) and
  \(\isotope[2]{H}(\isotope[10]{Be},\isotope[4]{He})\isotope[8]{Li}\) with
  44~\(A\)\,MeV \isotope[10]{Be} Radioactive Beam at ACCULINNA-2
  Fragment Separator}
\author {E.  Yu. Nikolskii}
\email{enikolskii@mail.ru}
\affiliation{National Research Centre ``Kurchatov Institute'', Moscow, Russia}
\affiliation{JINR, Flerov Laboratory of Nuclear Reactions, Dubna, Russia}
\author    {S.  A.  Krupko}
\affiliation{JINR, Flerov Laboratory of Nuclear Reactions, Dubna, Russia}
\author  {I.  A.  Muzalevskii}
\affiliation{JINR, Flerov Laboratory of Nuclear Reactions, Dubna, Russia}
\affiliation{Institute of Physics in Opava Silesian University in Opava, Opava, Czech Republic}
\author  {A.  A.  Bezbakh}
\affiliation{JINR, Flerov Laboratory of Nuclear Reactions, Dubna, Russia}
\affiliation{Institute of Physics in Opava Silesian University in Opava, Opava, Czech Republic}
\author    {R.      Wolski}
\affiliation{JINR, Flerov Laboratory of Nuclear Reactions, Dubna, Russia}
\author    {C.      Yuan}
\affiliation{Sino-French Institute of Nuclear Engineering and Technology (IFCEN),
  Sun Yat-sen University, Zhuhai, China}
\author  {S.  G.  Belogurov}
\affiliation{JINR, Flerov Laboratory of Nuclear Reactions, Dubna, Russia}
\affiliation{National Research Nuclear University ``MEPhI'', Moscow, Russia}
\author    {D.      Biare}
\affiliation{JINR, Flerov Laboratory of Nuclear Reactions, Dubna, Russia}
\author  {V.      Chudoba}
\affiliation{Institute of Physics in Opava Silesian University in Opava, Opava, Czech Republic}
\affiliation{JINR, Flerov Laboratory of Nuclear Reactions, Dubna, Russia}
\author  {A.  S.  Fomichev}
\affiliation{JINR, Flerov Laboratory of Nuclear Reactions, Dubna, Russia}
\affiliation{Dubna State University, Dubna, Russia}
\author    {E.  M.  Gazeeva}
\affiliation{JINR, Flerov Laboratory of Nuclear Reactions, Dubna, Russia}
\author    {M.  S.  Golovkov}
\affiliation{JINR, Flerov Laboratory of Nuclear Reactions, Dubna, Russia}
\author    {A.  V.  Gorshkov}
\affiliation{JINR, Flerov Laboratory of Nuclear Reactions, Dubna, Russia}
\author{L.  V.  Grigorenko}
\affiliation{JINR, Flerov Laboratory of Nuclear Reactions, Dubna, Russia}
\affiliation{National Research Centre ``Kurchatov Institute'', Moscow, Russia}
\affiliation{National Research Nuclear University ``MEPhI'', Moscow, Russia}
\author  {G.      Kaminski}
\affiliation{JINR, Flerov Laboratory of Nuclear Reactions, Dubna, Russia}
\author  {M.      Khirk}
\affiliation{JINR, Flerov Laboratory of Nuclear Reactions, Dubna, Russia}
\affiliation{Skobeltsyn Institute of Nuclear Physics, Moscow State University,%
  119991 Moscow, Russia}
\author    {O.      Kiselev}
\affiliation{GSI Helmholtzzentrum f\"ur Schwerionenforschung GmbH,
  Darmstadt, Germany}
\author {D.  A.  Kostyleva}
\affiliation{GSI Helmholtzzentrum f\"ur Schwerionenforschung GmbH,
  Darmstadt, Germany}
\affiliation{Physikalisches Institut, Justus-Liebig-Universit\"at,
  Giessen, Germany}
\author    {B.      Mauyey}
\affiliation{JINR, Flerov Laboratory of Nuclear Reactions, Dubna, Russia}
\affiliation{L.N. Gumilyov Eurasian National University, Astana, Kazakhstan}
\author    {I.      Mukha}
\affiliation{GSI Helmholtzzentrum f\"ur Schwerionenforschung GmbH,
  Darmstadt, Germany}
\author    {Yu. L.  Parfenova}
\affiliation{JINR, Flerov Laboratory of Nuclear Reactions, Dubna, Russia}
\author {A.  M.  Quynh}
\affiliation{JINR, Flerov Laboratory of Nuclear Reactions, Dubna, Russia}
\affiliation{Nuclear Research Institute, Dalat, Vietnam}

\author    {S.  I.  Sidorchuk}
\affiliation{JINR, Flerov Laboratory of Nuclear Reactions, Dubna, Russia}
\affiliation{Voronezh state university, Voronezh, Russia}
\author  {P.~G.  Sharov}
\email{sharovpavel@jinr.ru}
\affiliation{JINR, Flerov Laboratory of Nuclear Reactions, Dubna, Russia}
\affiliation{Institute of Physics in Opava Silesian University in Opava, Opava, Czech Republic}
\author  {N.~B.  Shulgina}
\affiliation{National Research Centre ``Kurchatov Institute'', Moscow, Russia}
\affiliation{Bogoliubov Laboratory of Theoretical Physics, JINR,  Dubna, Russia}
\author    {R.  S.  Slepnev}
\affiliation{JINR, Flerov Laboratory of Nuclear Reactions, Dubna, Russia}
\author    {S.  V.  Stepantsov}
\affiliation{JINR, Flerov Laboratory of Nuclear Reactions, Dubna, Russia}
\author {A.      Swiercz}
\affiliation{JINR, Flerov Laboratory of Nuclear Reactions, Dubna, Russia}
\author  {G.  M.  Ter-Akopian}
\affiliation{JINR, Flerov Laboratory of Nuclear Reactions, Dubna, Russia}
\affiliation{Dubna State University, Dubna, Russia}

\begin{abstract}
  The proton and deuteron pickup reactions
  \(\isotope[2]{H}(\isotope[10]{Be},\isotope[3]{He})\isotope[9]{Li}\) and
  \(\isotope[2]{H}(\isotope[10]{Be},\isotope[4]{He})\isotope[8]{Li}\)
  were studied with radioactive beam produced by the new fragment separator
  ACCULINNA-2 at   FLNR, JINR\@.
  These measurements were initially motivated as the test reactions intended
  for the elucidation of results obtained in the study of the extremely
  neutron-rich \isotope[7]{H} and \isotope[6]{H} systems created in the
  \(\isotope[2]{H}(\isotope[8]{He},\isotope[3]{He})\isotope[7]{H}\) and
  \(\isotope[2]{H}(\isotope[8]{He},\isotope[4]{He})\isotope[6]{H}\)
  reactions with the use of the same setup.
  In the \(\isotope[2]{H}(\isotope[10]{Be},\isotope[3]{He})\isotope[9]{Li}\)
  reaction the \isotope[9]{Li} ground-state (\(3/2^-\)) and its first excited
  state (2.69~MeV, \(1/2^-\)) were identified in the low-energy region of its
  excitation spectrum.
  The differential cross sections for the \(\isotope[9]{Li}_{\text{g.s.}}\)
  population were extracted at the forward center-of-mass angles
  (\(3^\circ\)--\(13^\circ\)) and compared with the FRESCO calculations.
  Spectroscopic factor of \(\sim\)1.7, derived by a model suggesting the
  \(\isotope[10]{Be} = p + \isotope[9]{Li}_{\text{g.s.}}\) clustering was found
  in accord with the experimental data.
  The energy spectrum of \isotope[8]{Li} populated in the
  \(\isotope[2]{H}(\isotope[10]{Be},\isotope[4]{He})\isotope[8]{Li}\)
  reaction shows the strong peak which corresponds to the excitation of the
  second excited state of \isotope[8]{Li} (2.25 MeV, \(3^+\)).
  The fact that the ground and the first excited states of \isotope[8]{Li} were
  not observed in this reaction is consistent with the shell-model structure of
  the nuclei involved.

\end{abstract}
\maketitle

\section{Introduction}
When conducting multi-parameter experiments dedicated to the
study of neutron drip-line nuclei obtained with very low yield in nuclear
reactions, the choice of reference reaction(s) is an important issue.
Such additional measurements are necessary in order to check: 
(i) whether the particle detectors work correctly,
(ii) how well the carried out Monte-Carlo (MC) simulations reflect
the real experimental conditions and
(iii) how accurately are estimated the main characteristics of
the experiment, such as the spectrum calibration and resolution assumed over
the measured excitation energy of the studied nucleus,
the detection efficiency of reaction products, and so on.
All together it can provide a new reliable information
(level scheme, spin--parity data, decay channels etc.)
for the unknown exotic systems.

Recently, the \isotope[7]{H} and \isotope[6]{H} systems were studied in
experiments~\cite{Bezbakh:2020,Muzalevskii:2021,Nikolskii:2022}
carried out at the fragment separator ACCULINNA-2 in the
Flerov Laboratory of Nuclear Reactions (FLNR, JINR)~\cite{Fomichev:2018}.
The achieved intensity of the \isotope[8]{He} beam up to \(\approx\!\!\!10^5\) pps
and the corresponding detection systems allowed to perform new exclusive
measurements aimed, in particular, at the search for \isotope[7]{H}
resonances populated in the proton pick-up reaction
\(\isotope[2]{H}(\isotope[8]{He},\isotope[3]{He})\isotope[7]{H}\)%
~\cite{Bezbakh:2020,Muzalevskii:2021}.
Data collected in the second experiment~\cite{Muzalevskii:2021} allowed one
to derive information on the deuteron-pickup reaction
\(\isotope[2]{H}(\isotope[8]{He},\isotope[4]{He})\isotope[6]{H}\).
This reaction was analyzed in \cite{Nikolskii:2022} giving new important results
concerning the \isotope[6]{H} excited states.

To check the reliability of the new results obtained for the \isotope[7]{H} and
\isotope[6]{H} systems~\cite{Bezbakh:2020,Muzalevskii:2021,Nikolskii:2022},
it was decided to perform reference measurements using the \isotope[10]{Be}
secondary beam and thereby to measure the same proton and deuteron pickup
reactions, i.~e.
\(\isotope[2]{H}(\isotope[10]{Be},\isotope[3]{He})\isotope[9]{Li}\) and
\(\isotope[2]{H}(\isotope[10]{Be},\isotope[4]{He})\isotope[8]{Li}\).
Originally, these measurements assumed the two main goals:
(i) to check the setup calibration over the excitation energy range of
\isotope[7]{H} and \isotope[6]{H}, and
(ii) to determine the real experimental excitation energy resolution which
should be compared with the Monte-Carlo calculations.

At the same time it turns out that the study of the
\(\isotope[2]{H}(\isotope[10]{Be},\isotope[3]{He})\isotope[9]{Li}\) and
\(\isotope[2]{H}(\isotope[10]{Be},\isotope[4]{He})\isotope[8]{Li}\) reactions
is in itself of considerable interest.
In particular, by measuring the
\(\isotope[2]{H}(\isotope[10]{Be},\isotope[3]{He})\isotope[9]{Li}_{\text{g.s.}}\)
reaction channel, the spectroscopic factor (SF) of the ground state of the
\isotope[10]{Be} nucleus, representing the system
\(\isotope[10]{Be} = \isotope[9]{Li}_{\text{g.s.}} + p\), could be deduced.
Accordingly, this result had stimulated corresponding theoretical calculations
which were compared with experiment.
Note that only one experiment was known before where the \isotope[9]{Li} nucleus
was studied in the \(d + \isotope[10]{Be}\) collisions~\cite{Kashy:1975}.
In this work only information about the ground state mass of \isotope[9]{Li}
was derived.
The performed for the first time spectroscopic study of \isotope[8]{Li} nucleus,
being the product of the
\(\isotope[2]{H}(\isotope[10]{Be},\isotope[4]{He})\isotope[8]{Li}\)
reaction allowed us to obtain information about the \isotope[10]{Be}
ground state and the \isotope[8]{Li} level structure that could be compared with
Shell-Model predictions.

\section{Experimental Setup}

The experiment was carried out at the Flerov Laboratory of Nuclear Reactions (JINR)
at the ACCULINNA-2 fragment separator~\cite{Fomichev:2018}.
The primary beam of \isotope[15]{N} ions with energy 49.7 \(A\)\*MeV and
intensity \(\approx\!\! 0.5\)~p\(\mu\)A, accelerated by the U-400M cyclotron,
bombarded the 1-mm thick beryllium production target installed at the initial
focal plane of the separator.
The secondary \isotope[10]{Be} radioactive ion beam (RIB) was obtained at
the final focal plane of the separator where the cryogenic deutrium
gas target was installed.
The \isotope[10]{Be} beam characteristics were as follows:
intensity \(\approx\!\! 10^5\)~pps, 80\% purity, with energy in the middle of
the \(D_2\) target 42~\(A\)\*MeV and energy spread \(\pm 2.5\%\).

\begin{figure}
  \setcaptionmargin{5mm}
  \onelinecaptionsfalse
  \includegraphics{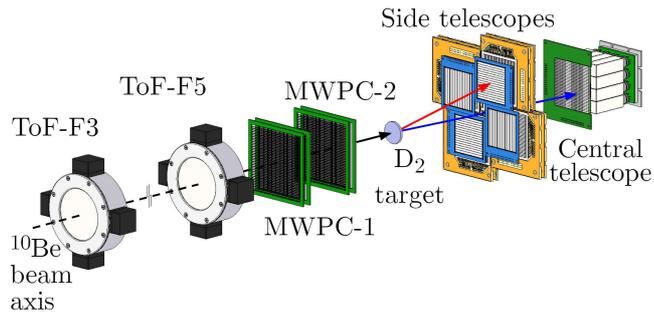}
  \captionstyle{normal} \caption{\label{fig:exp-setup}
    Experimental setup to study the
    \(\isotope[2]{H}(\isotope[10]{Be},\isotope[3]{He})\isotope[9]{Li}\) and
    \(\isotope[2]{H}(\isotope[10]{Be},\isotope[4]{He})\isotope[8]{Li}\)
    reactions with \isotope[10]{Be} secondary beam  at the ACCULINNA-2
    fragment separator.
    The time-of-flight scintillation sensors (ToF-F3, ToF-F5),
    position-sensitive beam detectors (MWPC-1,MWPC-2) are installed upstream
    the \(D_2\) cryogenic gas target.
    Charged particle detector telescopes for the
    \isotope[3]{He}, \isotope[4]{He} recoils and \isotope[7,8]{Li}
    decay products are placed downstream the \(D_2\) target.}
\end{figure}

\begin{figure}
  \setcaptionmargin{5mm}
  \onelinecaptionsfalse
  \includegraphics{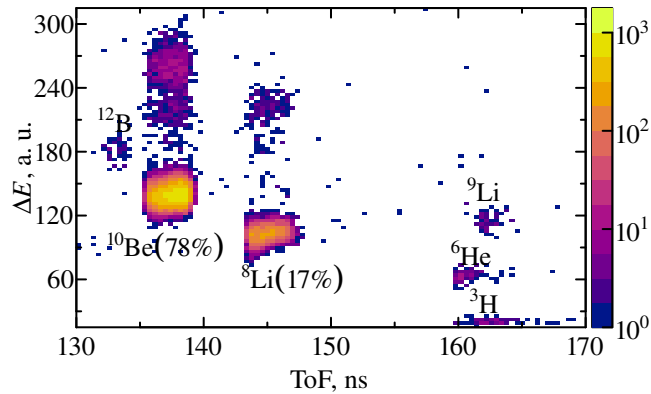}
  \captionstyle{normal} \caption{\label{fig:beam-ID}
    Particle identification plot measured at the ACCULINNA-2 F5 focal plane
    when the setup was tuned to the \isotope[10]{Be} secondary beam.
    The vertical axis corresponds to the energy loss in the ToF F5
    scintillator while the time of flight is plotted on the horizontal axis.}
\end{figure}

The experimental setup shown in Fig.~\ref{fig:exp-setup} was the same as
in~\cite{Bezbakh:2020,Muzalevskii:2021}.
Two BC-404 plastic scintillators placed at the F3 and F5 focal planes at the
time-of-flight (ToF) base 12.3~m allowed one  to identify the RIB projectiles
by the \(dE\)--ToF method~\cite{Kaminski:2020} and determine their energy with
precision \(\approx\!\! 0.5\%\)~(FWHM).
In Fig.~\ref{fig:beam-ID} we demonstrate the particle identification of RIB
obtained at the F5 focus.
It is seen that the \isotope[10]{Be} projectiles are well separated from
neighboring \isotope[8]{Li} and \isotope[12]{B} nuclei.
The low-intensity admixture of isotopes \isotope[3]{H}, \isotope[6]{He} and
\isotope[9]{Li} with \(A/Q = 3\) at ToF = 160--165 ns can be observed and well
separated too.
The trajectories of individual projectiles were measured by a pair of
multi-wire proportional chambers (MWPCs) placed at the distances 28 and 81 cm
upstream of the \(D_2\) target.
This allowed one to measure the coordinates of the interaction points in
the target plane with a 1.8-mm accuracy as well as to determine the inclination
angles of each projectile to the ion-optical axis with precision
of \(\approx\!\! 3\) mrad.

The cryogenic target gas cell was supplied with the 6-\(\mu\)m-thick and
25-mm diameter stainless-steel entrance and exit windows and filled with
deuterium gas at 27 K. The gas pressure was 1.13 atm. (``thick'' mode) or
0.56 atm (``thin'' mode) resulted in target thickness \(3.7 \times 10^{20}\) or
\(1.8 \times 10^{20}\) deuterons/cm\(^2\), respectively.
The gas cell was covered by a pair of 3.5-\(\mu\)m aluminum-backed Mylar
screens kept cooled to the same temperature to ensure thermal protection.

The main contribution to the \isotope[9]{Li} (\isotope[8]{Li}) excitation energy
resolution was made by the energy determination of
the \isotope[3]{He} (\isotope[4]{He}) recoils emitted in the studied reaction.
The error in the measured energy was mainly caused by the uncertainty of the
interaction point over ``Z'' coordinate (along the beam axis)
in the target volume.
It was assumed in the data analysis that the interaction point was in
the middle plane of the target.
To ensure a homogeneous thickness of the target, only events generated within
the circle in the target cell with diameter 17 mm were taken into account.
This selection also ensured the rejection of the reactions with the material
of the target frame.

\begin{figure}
  \setcaptionmargin{5mm}
  \onelinecaptionsfalse
  \includegraphics{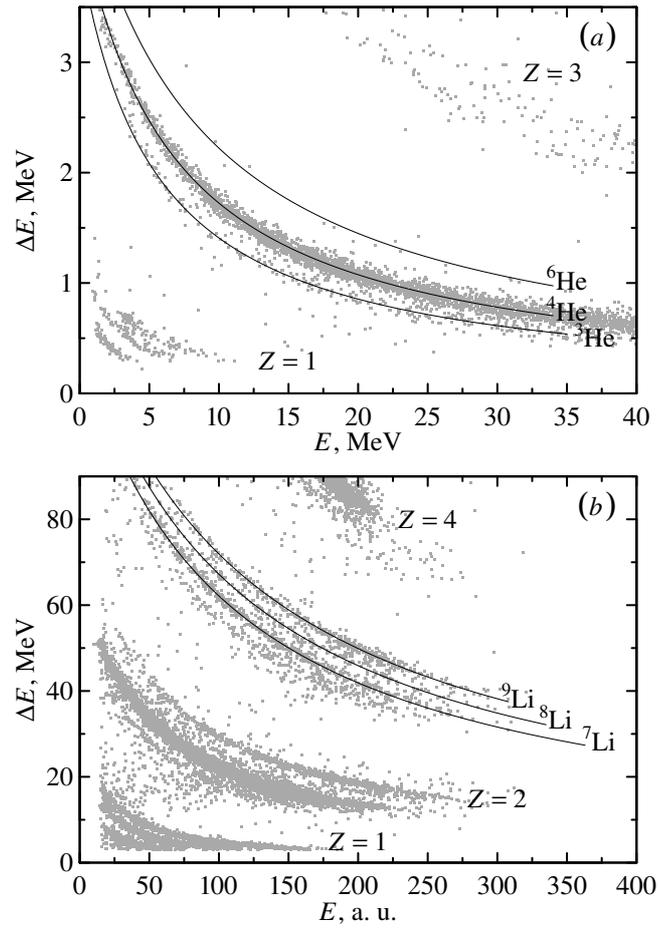}
  \captionstyle{normal} \caption{\label{fig:dEE-ID}
    (\textit{a}) \(\Delta E\)--\(E\) particle identification plot obtained from
    the 16 strips of one of the recoil telescopes after the thickness correction.
    The solid lines show the locations of  \isotope{He}-isotopes.
    It is seen that the \isotope[3]{He} and \isotope[4]{He} recoil nuclei are
    clearly separated;
    (\textit{b}) The same plot obtained in central telescope;
    The solid lines show the locations of  \isotope{Li}-isotopes.
  }
\end{figure}

The telescope assembly (side telescopes, see Fig.~\ref{fig:exp-setup})
intended to detect \isotope[3,4]{He}-recoils was
installed at a distance of 179 mm from the target.
It consisted of the four identical \(dE\)--\(E\)--\(E\) telescopes made of
the single-sided silicon detectors (SSD).
The angular range covered by these telescopes for the \isotope[3,4]{He} recoil
nuclei extended from \(\approx\!\! 6^\circ\) to \(\approx\!\! 24^\circ\) in
the laboratory system.
The 20-\(\mu\)m thick SSD with a sensitive area \(50 \times 50\) mm\(^2\) was
divided into 16 strips, the second and the third telescope layers were created
by the two identical 1-mm thick SSDs
(each with \(60 \times 60\) mm\(^2\) area with segmentation in 16 strips),
where the last \(E\)-detector was used as the Veto-sensor.
To provide particle identification, a sophisticated procedure was developed for
the \isotope{He}-recoil telescopes.
Since the energy of the recoils leaving the target was expected to be too low
(starts from \(\approx\!\! 7\) MeV), very thin, 20-\(\mu\)m \(\Delta E\)-detectors
were needed for the \isotope[3,4]{He} identification.
Because of the fabrication inhomogeneity, inherent to the so thin
silicon wafers, the thickness maps were determined by calibration for each
of these thin detectors~\cite{Muzalevskii:2020}.
Fig.~\ref{fig:dEE-ID}\textit{a} illustrates the particle identification obtained for
all strips in one thin detector, which was actually implemented for each strip
separately, but even in this presentation it looks quite convincing.

The lithium isotopes emitted as a result of \(\isotope[10]{Be}+\isotope[2]{H}\)
collisions were detected by the \(64 \times 64~\text{mm}^2\) 
central telescope installed at zero laboratory angle 323 mm downstream
the target, see Fig.~\ref{fig:exp-setup}.
The telescope consisted of one 1500-micron Si double-side detector
(\(32 \times 32\) strips with 2 mm pitch)  
and a set of 16 square CsI(Tl)/PMT modules.
The CsI(Tl) crystals had a cross section of \(16.5 \times 16.5~\text{mm}^2\) and
thickness 50 mm each, which allowed to stop all charged particles in
the sensitive volume of the telescope.
Each crystal was covered with a 3.5-\(\mu\)m-thick aluminized Mylar on its
entrance and was coupled with the Hamamatsu R9880U-20 photomultiplier tube (PMT)
by the optical grease.
The particle identification plot obtained in the central telescope is shown in
Fig.~\ref{fig:dEE-ID}\textit{b}.

The key issues of these measurements were
(i) the clear \isotope[3]{He} and \isotope[4]{He} identification provided at the
low particle energy (\(\sim\)7--20 MeV),
(ii) the good energy and angular resolution of the experimental setup, and
(iii) the suficiently high efficiency of coincidence events occurring between
the \isotope[3,4]{He} recoils and the \isotope[7,8]{Li} nuclei emitted at
forward direction.

\section{\(\isotope[2]{H}(\isotope[10]{Be},\isotope[3]{He})\isotope[9]{Li}\)
  Reaction}
\begin{figure}
  \centering
  \setcaptionmargin{5mm}
  \onelinecaptionsfalse
  \includegraphics{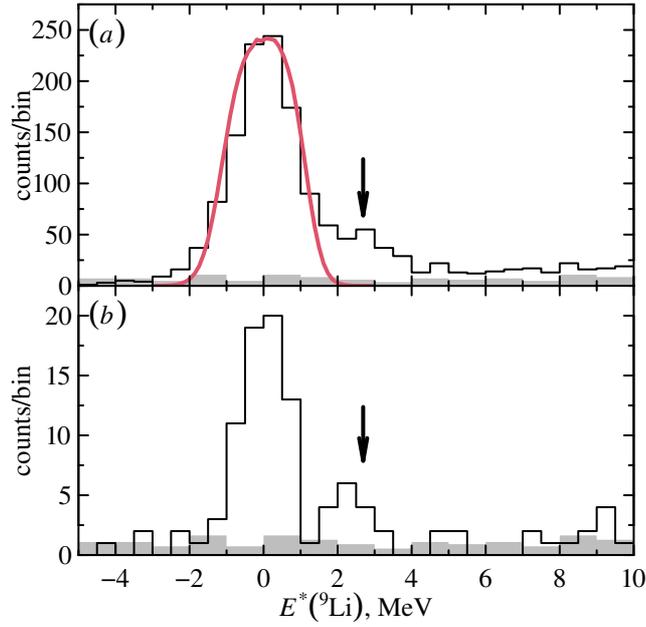}
  \captionstyle{normal} \caption{\label{fig:9Li-spectra}
    The excitation energy spectra of \isotope[9]{Li} measured in the
    \(\isotope[2]{H}(\isotope[10]{Be}, \isotope[3]{He})\isotope[9]{Li}\)
    reaction with the  ``thick'' (\textit{a}) and ``thin'' (\textit{b}) \(D_2\) gas targets.
    The solid curve in panel (\textit{a}) represents the MC calculation of the
    \(\isotope[9]{Li}_{\text{g.s.}}\) taking the parameters of
    the experimental setup.
    The gray filled histograms represent the background level measured with the
    empty target and normalized on the \isotope[10]{Be} beam integral.
    The arrows mark \isotope[9]{Li} first excited state with \(J^\pi=1/2^-\) and
    \(E^*=2.69\)~MeV.  }
\end{figure}

The excitation spectra of \isotope[9]{Li} nucleus derived from the data obtained
in the runs with \isotope[10]{Be} bombarding the ``thick''  and ``thin'' D\(_2\)
target are presented in Figs.~\ref{fig:9Li-spectra}\textit{a} and
\ref{fig:9Li-spectra}\textit{b}, respectively.
The spectra are obtained by the missing mass (MM) method realized by taking the
recorded event-by-event data on the \isotope[10]{Be} projectiles
(the energy, the trajectory angle, and hit position on the target)
and corresponding \isotope[3]{He} recoils
(the measured energy and emission angle).
The black solid-line histogram in Fig.~\ref{fig:9Li-spectra}\textit{a} shows
a well-pronounced peak corresponding to the ground state of \isotope[9]{Li}
populated in the
\(\isotope[2]{H}(\isotope[10]{Be},\isotope[3]{He})\isotope[9]{Li}_\text{g.s.}\)
reaction.
On the right slope of this peak, the presence of the not well-resolved
first excited state of \isotope[9]{Li} (\(E^* = 2.69\) MeV~\cite{Tilley:2004})
is also observed.
The solid curve in Fig.~\ref{fig:9Li-spectra}\textit{a} demonstrates
the Monte-Carlo calculation for the \(\isotope[9]{Li}_\text{g.s.}\) using
the given above parameters of the experimental setup.

One can clearly see that the MC simulation reproduces quite well the shape of
the \(\isotope[9]{Li}_\text{g.s.}\) peak demonstrating experimental resolution
\(\approx\!\! 2.2\) MeV.
Note that in experiment~\cite{Muzalevskii:2021} the corresponding calculations
of the resolution of \isotope[7]{H} populated in the
\(\isotope[2]{H}(\isotope[8]{He},\isotope[3]{He})\isotope[7]{H}\) reaction
at energy near 2 MeV gave the much better value of \(\approx\!\! 1.1\) MeV.
The reason for this \(\approx\!\! 2\) times better resolution in the \isotope[7]{H}
experiment is caused by the larger energy of the \isotope[3]{He} recoils,
as compared to the
\(\isotope[2]{H}(\isotope[10]{Be},\isotope[3]{He})\isotope[9]{Li}\) reaction,
and, therefore, the smaller energy losses in the target.
It shows that the target thickness makes the main contribution to the energy
resolution in this energy range.
This is clearly seen on Fig.~\ref{fig:9Li-spectra}\textit{b} where
the measurement on ``thin'' target is presented.
One can see that the spectrum, obtained on ``thin'' target, demonstrates a good
separation between the ground and the first excited (2.69 MeV) states
of \isotope[9]{Li}.

\begin{figure}
  \setcaptionmargin{5mm}
  \onelinecaptionsfalse
  \includegraphics{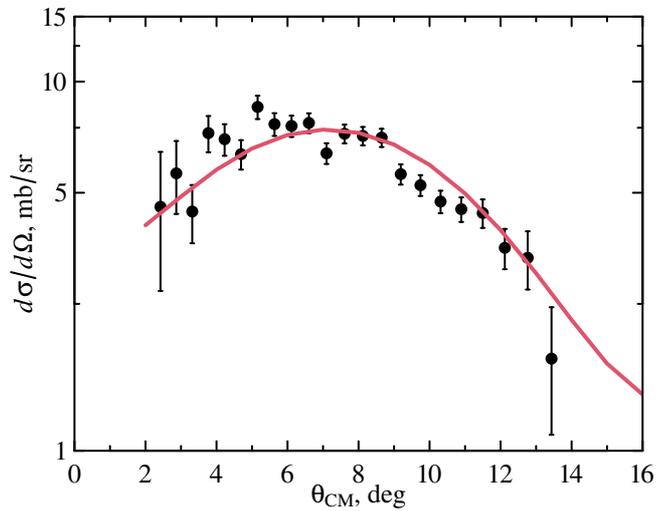}
  \captionstyle{normal} \caption{\label{fig:9Li-ang-distr}
    The differential cross-sections for the
    \(\isotope[2]{H}(\isotope[10]{Be},\isotope[3]{He})\isotope[9]{Li}\) reaction.
    Points show
    the angular distribution  for the
    \(\isotope[2]{H}(\isotope[10]{Be},\isotope[3]{He})\isotope[9]{Li}_\text{g.s.}\)
    reaction obtained in the experiment. 
    (The error bars are statistical only.)
    The solid curve shows the result of corresponding
    FRESCO one-step calculations, see text for the details.
  }
\end{figure}

Fig.~\ref{fig:9Li-ang-distr} shows the angular distribution extracted from the
experimental data at the forward center-of-mass angles (\(3^\circ\)--\(13^\circ\))
for the \isotope[9]{Li} ground state populated in the
\(\isotope[2]{H}(\isotope[10]{Be},\isotope[3]{He})\isotope[9]{Li}\) reaction. 

The differential cross sections were analyzed with the Coupled Reaction Channels
Calculations approach using the code FRESCO~\cite{Thompson:1988}.
For the entrance (\(d + \isotope[10]{Be}\)) and exit
(\(\isotope[3]{He} + \isotope[9]{Li}\)) channels, the Global Wood-Saxon
optical model potentials found in Ref.~\cite{Daehnick:1980} and
in Ref.~\cite{Pang:2015}, respectively, were used.
In the calculation, spectroscopic factor (SF) for the \(\isotope[3]{He}=p\,+\,d\)
clustering was set as SF = 1.32, according to the \emph{ab initio} value
obtained by I.~Brida et al. \cite{Brida:2011}.

It should be noted that before our research, the SF for the system
\(\isotope[10]{Be}_{\text{g.s.}} = p + \isotope[9]{Li}(3/2^-)\) was unknown.
To evaluate the corresponding SF we carried out the shell-model calculation
KSHELL~\cite{Shimizu:2019} with the modified version of ``YSOX'' Hamiltonian
based on the monopole universal interaction~\cite{Yuan:2012}.
Model space was \(psd\), including five orbits \(0p_{3/2}\), \(0p_{1/2}\),
\(0d_{5/2}\), \(0d_{3/2}\), \(1s_{1/2}\) for both, protons and neutrons.
Excitation from \(p\)- to \(sd\)-shell was limited to 0--5 nucleons.
This (0--5)\(\hbar\omega\) constraint just was chosen just for the consistency
comparing with calculations done for the C, N, O isotopes~\cite{Yuan:2012}
where the Hamiltonian can reproduce well the ground-state energies, drip lines,
energy levels, electric properties, and spin properties of \(psd\)-shell nuclei.
Table~\ref{tab:SF-9Li-p} presents the SF calculations of 
\(\isotope[10]{Be}\to \isotope[9]{Li} + p\) clustering for the
\(\isotope[10]{Be}(0^+)\) ground and first excited \(\isotope[10]{Be}(2^+)\)
states.

Finally, with this Hamiltonian, SF = 1.74 was obtained for the
\(\isotope[10]{Be}_{\text{g.s.}} = p + \isotope[9]{Li}(3/2^-)\) clustering while
the \(p + \isotope[9]{Li}(1/2^-)\) cluster configuration was found to be much
less, SF = 0.207.
As it is seen from Fig.~\ref{fig:9Li-ang-distr}, the solid curve,
which corresponds to the FRESCO calculation with the mentioned SFs for the
\(\isotope[10]{Be}_{\text{g.s.}}\) clustering, demonstrates
the consistency of this model with the experimental data.
Thus, the measured differential cross sections of the
\(\isotope[2]{H}(\isotope[10]{Be},\isotope[3]{He})\isotope[9]{Li}_{\text{g.s.}}\)
reaction at forward center-of-mass angles confirm the modern theoretical
approach made for the ground-state structure of the \isotope[10]{Be} nucleus.

\section{\(\isotope[2]{H}(\isotope[10]{Be},\isotope[4]{He})\isotope[8]{Li}\)
  Reaction}

\begin{figure}
  \setcaptionmargin{5mm}
  \onelinecaptionsfalse
  \includegraphics{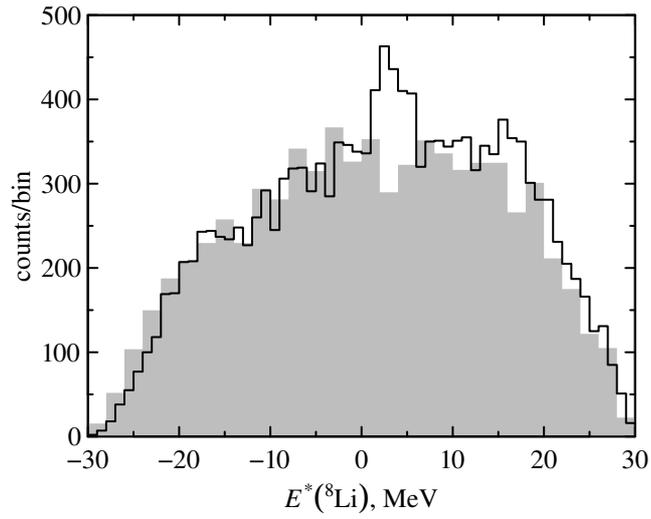}
  \captionstyle{normal} \caption{\label{fig:8Li-spectra}
    The inclusive MM spectrum of \isotope[8]{Li} from the
    \(\isotope[2]{H}(\isotope[10]{Be},\isotope[4]{He})\isotope[8]{Li}\)
    reaction.
    The gray filled histogram corresponds to the background measurement with
    the empty target normalized on the beam integral.
  }
\end{figure}

In the same run, the
\(\isotope[2]{H}(\isotope[10]{Be},\isotope[4]{He})\isotope[8]{Li}\) reaction
was studied by detecting the \isotope[4]{He} recoils.
The inclusive missing mass spectrum of \isotope[8]{Li} obtained in the four
recoil telescopes is shown in Fig.~\ref{fig:8Li-spectra}.
One can see a peak at around \(E^*(\isotope[8]{Li}) =2.5\)~MeV sitting on
a large background.
The nature of this background is mainly associated with nuclear reactions
generated by the \isotope[10]{Be} projectiles on the entrance and exit windows
of the target cell.
We made special background measurements without D\(_2\) gas in the target cell.
This background is shown by the filled histogram in Fig.~\ref{fig:8Li-spectra}.
Note that the \isotope[4]{He}-recoil background level is much higher than that
in the case of \isotope[3]{He}.
This situation was quite expected since the yield of alpha particles is high
because of the intense reaction channels in which the formation of slow alpha
particles occurs with a much larger cross section than for the other
light nuclei (cluster effects, preferred energy dependence, etc.)

\begin{figure}
  \setcaptionmargin{5mm}
  \onelinecaptionsfalse
  \includegraphics{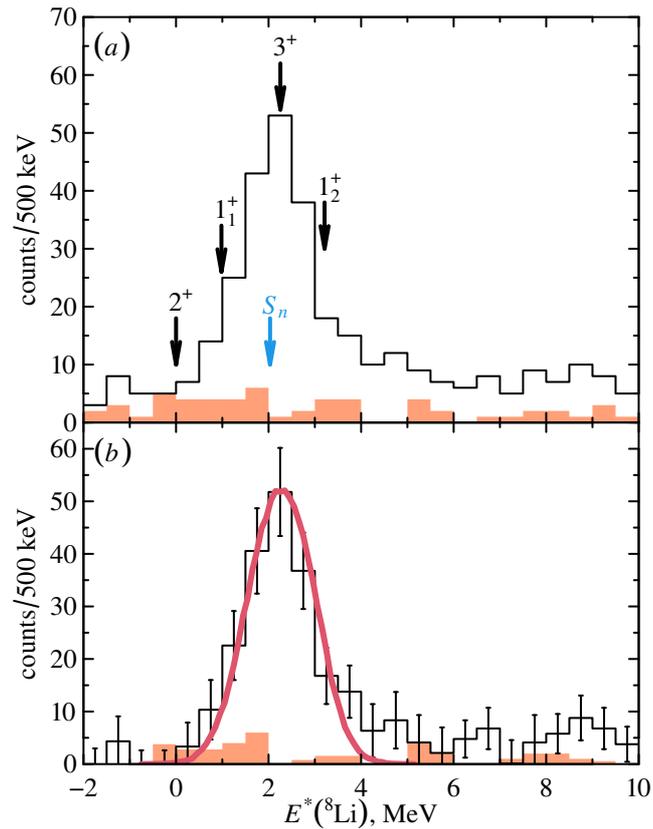}
  \captionstyle{normal} \caption{\label{fig:8Li-spectra-2}
    (\textit{a}) The excitation spectra of \isotope[8]{Li} obtained in the case
    of the \isotope[4]{He} recoil in coincidence with \isotope[8]{Li}
    (filled histogram) and \isotope[7]{Li} (black histogram).
    The arrows mark the excitation energy for the ground (\(2^+\)), first (\(1^+_1\)),
    second (\(3^+\)), third (\(1^+_2\)) \isotope[8]{Li} excited states, and
    the \isotope[8]{Li} neutron separation energy (\(S_n\)).
    (\textit{b}) The excitation spectrum of \isotope[8]{Li} obtained from the
    \(\isotope[2]{H}(\isotope[10]{Be},\isotope[4]{He})\isotope[8]{Li}\)
    reaction data after the empty-target subtraction with the condition of
    coincidences \isotope[4]{He}--\isotope[7]{Li} (black histogram)
    and \isotope[4]{He}--\isotope[8]{Li} (filled histogram).
    The solid curve corresponds to the Monte-Carlo calculation for the
    contribution of the 2.255 MeV (\(3^+\)) \isotope[8]{Li} state.
  }
\end{figure}

In order to reduce the background we applied the condition of coincidences with
the \isotope{Li} isotopes detected in the central telescope.
The obtained spectra are shown in Fig.~\ref{fig:8Li-spectra-2}\textit{a}.
As it is seen from Fig.~\ref{fig:8Li-spectra-2}~(a), the coincidences with
\isotope{Li} nuclei drastically reduced the background, and now the structure at
\(E^* \sim 2.5\)~MeV is clearly manifested.
In addition, it is important to note that almost all events near
the 2.5-MeV peak are exhausted by the coincidences with \isotope[7]{Li}, which
are shown by black histogram in Fig.~\ref{fig:8Li-spectra-2}~(a).
It means that the 2.5-MeV structure is associated with the states of
\isotope[8]{Li} decaying by neutron emission into \isotope[7]{Li}.
Since the ground (\(2^+\)) and the first excited (0.980 MeV, \(1^+\)) states of
\isotope[8]{Li} are nucleon stable
(see the energy level list of \isotope[8]{Li} in Table~\ref{tab:8Li-lev-scm}),
the peak near 2.5 MeV can be explained by the contribution of the 2.255 MeV
(\(3^+\)) state with possible weak input of the 3.21 MeV (\(1^+\)) state.

Then, the question naturally arises: why the ground and first excited states of
\isotope[8]{Li} are not populated in this reaction?

We considered this question taking the known theoretical shell-model calculations
made for the \(\isotope[10]{Be}_{\text{g.s.}}\) and \isotope[8]{Li}
level structures and then applied the quantum-number selection rules.
The followings issues should be taken into account: according to the Boyarkina
study of the \(p\)-shell nuclei~\cite{Boyarkina:1973} the \(\isotope[10]{Be}(0^+)\)
ground state is a multi-component ensemble of shell-model wave functions.
The most abundant one is $L=0$, \(S=0\), \(J=0\) component with weight 60\%,
the next two are of $L=1$, $S=1$, $J=0$ with smaller weights, 23\% and 11\%.
For the four \isotope[8]{Li} low-laying states only their largest components are
taken into account here.
The two initial levels of \isotope[8]{Li} are $L=1$, $S=1$ coupled to $J=2$ and
$J=1$, respectively~\cite{Boyarkina:1973}.
In the reaction discussed these levels can not be populated from the major
component of \(\isotope[10]{Be}_{\text{g.s.}}\) because of parity.
The non-zero excitation of these states could be due to the presence of
the minor \(\isotope[10]{Be}_{\text{g.s.}}\) components.
The second pair of \isotope[8]{Li} states are the members of $L=2$, $S=1$ multiplet
coupled to $J=3$ and $J=1$, respectively.
These states are excited stronger due to the \(\isotope[10]{Be}_{\text{g.s.}}\)
main component contribution to the transfer reaction transitions.
The component shell-model wave functions, shown in Table~\ref{tab:SA-8Li-d},
are characterized by \([f]^{ (2T+1)(2S+1)L}\), where \( [f]\) is the
Young scheme determining the permutational symmetry of the orbital part of the
wave function (w.~f.), $T$, $S$ and $L$ stand for the isospin, spin,
and orbital angular momentum, respectively, and their weights.
These w.~f.\ quantum numbers and weights are shown in the first and second rows,
respectively, for the three largest \(\isotope[10]{Be}_{\text{g.s.}}\)
components, and in the fourth and fifth columns for the dominant components of
the four \isotope[8]{Li} states.

The \(nlj\) in Table~\ref{tab:SA-8Li-d} are the quantum numbers of the transfer
transitions.
The spectroscopic amplitudes (SA) for the most significant
\(\isotope[10]{Be}_{\text{g.s.}}\to\isotope[8]{Li} + d\) partitions are
calculated in the frame of Translational Invariant Shell Model (TISM) applied to
\(p\)-shell nuclei~\cite{Smirnov:1977} by means of the genealogical
coefficients tables in~\cite{Neudachin:1969} or~\cite{Rudchik:1982}.
For completeness, SA for the \(\isotope[4]{He}\to d + d \)
clustering equal to 1.73 is taken from~\cite{Nemets:1988}.

The cross section of the
\(\isotope[2]{H}(\isotope[10]{Be},\isotope[4]{He})\isotope[8]{Li}\)
reaction leading to a particular \isotope[8]{Li} state is the square of
the coherent sum of the each component transition amplitudes.
It is seen in Table~\ref{tab:SA-8Li-d} that the contribution of the main
\(\isotope[10]{Be}_{\text{g.s.}}\) w.~f.\ component to the production of
the two initial \isotope[8]{Li} levels is parity forbidden.
Only the minor \(\isotope[10]{Be}_{\text{g.s.}}\) components participate in the
reaction.
Therefore, the resulting cross sections for these \isotope[8]{Li} states are
reduced in comparison to the higher-lying states of \isotope[8]{Li}.

The results of spectroscopic amplitude calculations shown in
Table~\ref{tab:SA-8Li-d} imply that the cross sections for the excitation of
the first two \isotope[8]{Li} states are reduced with respect to the
\(\isotope[8]{Li}(3^+)\) state, though they are not negligible.
Having in mind the experimental spectrum shown in
Fig.~\ref{fig:8Li-spectra-2}\textit{a} we conclude that our model predictions
overestimate the production the \(\isotope[8]{Li}_{\text{g.s.}}\) and
the \isotope[8]{Li} first excited state in the discussed reaction.

\begin{figure}
  \setcaptionmargin{5mm}
  \onelinecaptionsfalse
  \captionstyle{normal} \includegraphics{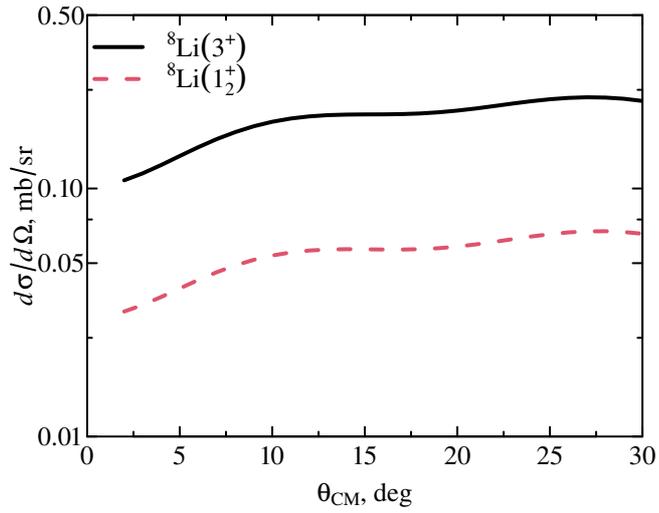}
  \caption{\label{fig:8Li-ang-distr}
    The FRESCO calculation of the differential cross sections for the
    \(\isotope[2]{H}(\isotope[10]{Be},\isotope[4]{He})\isotope[8]{Li}\)
    reaction populating the \(3^+\) state with \(E^*=2.255\)~MeV (solid line) and
    the \(1^+_2\) state  with \(E^*=3.21\) MeV (dashed line) of \isotope[8]{Li}
    (only pure main components of \(\isotope[10]{Be}_{\text{g.s.}}\) and
    \isotope[8]{Li} wave functions were taken into account). }
\end{figure}

We made FRESCO calculations of the differential cross sections for the
\(\isotope[2]{H}(\isotope[10]{Be},\isotope[4]{He})\isotope[8]{Li}\) reaction
leading to the population of 2.255 and 3.21 MeV states in \isotope[8]{Li},
see Fig.~\ref{fig:8Li-ang-distr}.
In the calculation, the spectroscopic factors for the configurations
\(\isotope[10]{Be}_{\text{g.s.}} = d  +  \isotope[8]{Li}(3^+)\)
and \(\isotope[10]{Be}_{\text{g.s.}} = d  + \isotope[8]{Li}(1^+)\) were taken as
0.54 and 0.16, respectively.
One can see that in the measured center-of-mass angular range
(\(\theta_{\text{CM}}\approx 4^\circ\text{--}16^\circ\))
the ratio of the cross sections obtained for the 2.25 and 3.21 MeV states
is about 4:1.

Finally, after the background subtraction, taking into account  empty target
measurement, we fitted the \isotope[8]{Li} spectrum using the  Monte-Carlo
response for the excited state 2.255 MeV (\(3^+\)).
The result is presented in Fig.~\ref{fig:8Li-spectra-2}\textit{b}.
It is seen that the spectrum is in a good agreement with the 2.255 MeV state.
However, we cannot rule out the small presence of 3.21 MeV state on the right
slope of the 2.255 peak.
This contribution around 15--20\% is in agreement with FRESCO 
estimation (see Fig.~\ref{fig:8Li-ang-distr}).
\section{Conclusions}
Excitation spectra of \isotope[9]{Li} and \isotope[8]{Li} obtained in the
\(\isotope[2]{H}(\isotope[10]{Be},\isotope[3]{He})\isotope[9]{Li}\) and
\(\isotope[2]{H}(\isotope[10]{Be},\isotope[4]{He})\isotope[8]{Li}\) reactions
were measured for the first time with the 44 \(A\)\*MeV \isotope[10]{Be}
radioactive beam produced by the fragment separator ACCULINNA-2 at the U-400M
cyclotron (FLNR, JINR).
The angular distribution was obtained for the
\(\isotope[2]{H}(\isotope[10]{Be},\isotope[3]{He})\isotope[9]{Li}_{\text{g.s.}}\)
reaction channel at the forward center-of-mass angles (\(3^\circ\)--\(13^\circ\)).
The absolute values of the cross sections are well predicted by the modern
shell-model calculation with spectroscopic factor SF \(\sim\!\! 1.7\) for the
\(\isotope[10]{Be} = p + \isotope[9]{Li}_{\text{g.s.}}\) configuration.
In the energy spectrum of \isotope[8]{Li} derived from the deuteron pickup reaction
\(\isotope[2]{H} (\isotope[10]{Be},\isotope[4]{He})\isotope[8]{Li}\),
a strong peak attributed to the second excited state of \isotope[8]{Li}
(2.25 MeV, \(3^+\)) was observed.
Also, the presence of an unresolved third excited state at 3.21 MeV (\(1^+\))
is possible in the measured \isotope[8]{Li} spectrum.
The absence of the ground and the first excited states of \isotope[8]{Li} in
the measured spectrum is explained in complete agreement with the theoretical
concepts of the structures of the \isotope[10]{Be} ground state and
\isotope[8]{Li} levels at low excitation energy.

In general, these and the previous
results~\cite{Bezbakh:2020,Muzalevskii:2021,Nikolskii:2022} of
the first experiments carried out with secondary beams at the ACCULINNA-2
facility show a great potential for the studies of light exotic nuclei
being close to and beyond the nuclear drip lines. 

\begin{acknowledgments}
This work was partly supported by the Russian Science Foundation grant
No.~22-12-00054.
One of the authors, I.~A.~M., was supported by the Student Grant Foundation of
the Silesian University in Opava, Grant No.~SGF/2/2020, which was realized
within the EU OPSRE project entitled
``Improving the quality of the internal grant scheme of
the Silesian University in Opava'',
No. CZ.02.2.69/0.0/0.0/19 073/00116951.
We acknowledge the interest and support of this activity from Prof.
Yu.Ts. Oganessian and Prof. S.N. Dmitriev.
\end{acknowledgments}
\bibliographystyle{maik}
\bibliography{9Li_8Li}

\clearpage 

\begin{table}
  \setcaptionmargin{0mm}
  \onelinecaptionsfalse
  \captionstyle{flushleft}
  \caption{\label{tab:SF-9Li-p}
    SFs for the \(\isotope[10]{Be}\to \isotope[9]{Li} + p\)
    clustering obtained by the Shell Model \cite{Shimizu:2019,Yuan:2012}.
  }
  \begin{ruledtabular}
    \begin{tabular}{llcc}
      Clustering & & \multicolumn{2}{c}{Transition}\\
        & & \(p_{1/2}\) &\(p_{3/2}\) \\\hline
      \(\isotope[10]{Be}_{\mathrm{g.\,s.}}=\isotope[9]{Li}(3/2^-)+p\)&& &1.74 \\
      \(\isotope[10]{Be}_{\mathrm{g.\,s.}}=\isotope[9]{Li}(1/2^-)+p\)&&0.207 &\\
    \(\isotope[10]{Be}(2^+)=\isotope[9]{Li}(3/2^-)+p\)&&0.156 &1.32\\
    \(\isotope[10]{Be}(2^+)=\isotope[9]{Li}(1/2^-)+p\)&& &0.184\\
    \(\isotope[10]{Be}(2^+)=\isotope[9]{Li}(5/2^-)+p\)&&0.0285 &0.0639\\
    \end{tabular}
  \end{ruledtabular}
\end{table}

\begin{table}
  \setcaptionmargin{0mm}
  \onelinecaptionstrue
  \captionstyle{flushleft}
  \caption{\label{tab:8Li-lev-scm}
    The list of low-energy \isotope[8]{Li} states based on~\cite{Tilley:2004}
  }
 \begin{ruledtabular}
   \begin{tabular}{cccc}
     \(E^*\), MeV &\(J^\pi\)&\(T_{1/2}/\Gamma\)& Decay mode\\
     \hline
     0.0          &\(2^+\) &839.9 ms       & \(\beta^-\)-decay\\
     0.9808       &\(1^+_1\) &8.2 fs & \(\gamma\)-emission\\
     2.255        &\(3^+\) &32.3 keV& \(n\)-emission\\
     3.210        &\(1^+_2\) & 1000 keV& \(n\)-emission\\
     5.4          &\(1^+_3\) & 650 keV & \(n\)-emission\\
    \end{tabular}
 \end{ruledtabular}
\end{table}

\begin{table*}
  \setcaptionmargin{0mm}
  \onelinecaptionstrue
  \captionstyle{flushleft}
  \caption{\label{tab:SA-8Li-d}
   Spectroscopic Amplitudes of the intense
   \(\isotope[10]{Be}_{\text{g.s.}}\to \isotope[8]{Li} + d\) clustering;
   the \(E^*\) given in MeV}
 \begin{ruledtabular}
   \begin{tabular}{rccccccc}
   &\(E^*\)&\(J^\pi\)&\({[f]}^{(2T+1)(2S+1)}L\)&&\([42]^{31}S\)&\([411]^{33}P\)&\([33]^{33}P\)\\
   \isotope[10]{Be} &0&\(0^+\)& &Weight&60\%&23\%&11\%\\
   \hline
          && & & & \multicolumn{3}{c}{\(nlj\) transition (Spectroscopic Amplitude) }\\
  \hline
   \isotope[8]{Li}&0&\(2^+\)&\([31]^{33}P\)&90.4\%&0.0&\(1D_2(0.293)\)& \(1D_2(0.102)\)\\
   \isotope[8]{Li}&0.980&\(1^+\)&\([31]^{33}P\)&78.5\%&0.0&\(2S_1(0.088)\)&\(2S_1(-0.245)\)\\
          &&&&&&\(1D_1(0.158)\)&\(1D_1(0.055)\)\\
   \isotope[8]{Li}&2.255&\(3^+\)&\([31]^{33}D\)&82.1\%&\(1D_3(-0.732)\)&&\\
   \isotope[8]{Li}&3.21&\(1^+\)&\([31]^{33}D\)&85.7\%&\(1D_1(-0.40)\)&&\\ 
   \isotope[4]{He} & ---&\(1^+\)& \([00]^{13}S\)&100\%&\(1S_1(1.73) \)&&\\ 
   \end{tabular}
 \end{ruledtabular}
\end{table*}
\clearpage 
\begin{enumerate}
  
\item     Experimental setup to study the
  \(\isotope[2]{H}(\isotope[10]{Be},\isotope[3]{He})\isotope[9]{Li}\) and
  \(\isotope[2]{H}(\isotope[10]{Be},\isotope[4]{He})\isotope[8]{Li}\)
  reactions with \isotope[10]{Be} secondary beam  at the ACCULINNA-2
  fragment separator.
  The time-of-flight scintillation sensors (ToF-F3, ToF-F5),
  position-sensitive beam detectors (MWPC-1,MWPC-2) are installed upstream
  the \(D_2\) cryogenic gas target.
  Charged particle detector telescopes for the
  \isotope[3]{He}, \isotope[4]{He} recoils and \isotope[7,8]{Li}
  decay products are placed downstream the \(D_2\) target.
\item     Particle identification plot measured at the ACCULINNA-2 F5 focal plane
  when the setup was tuned to the \isotope[10]{Be} secondary beam.
  The vertical axis corresponds to the energy loss in the ToF F5
  scintillator while the time of flight is plotted on the horizontal axis.
\item (\textit{a}) \(\Delta E\)--\(E\) particle identification plot obtained from
  the 16 strips of one of the recoil telescopes after the thickness correction.
  The solid lines show the locations of  \isotope{He}-isotopes.
  It is seen that the \isotope[3]{He} and \isotope[4]{He} recoil nuclei are
  clearly separated;
  (\textit{b}) The same plot obtained in central telescope;
  The solid lines show the locations of  \isotope{Li}-isotopes.
\item The excitation energy spectra of \isotope[9]{Li} measured in the
  \(\isotope[2]{H}(\isotope[10]{Be}, \isotope[3]{He})\isotope[9]{Li}\)
  reaction with the  ``thick'' (\textit{a}) and ``thin'' (\textit{b}) \(D_2\) gas targets.
  The solid curve in panel (\textit{a}) represents the MC calculation of the
  \(\isotope[9]{Li}_{\text{g.s.}}\) taking the parameters of
  the experimental setup.
  The gray filled histograms represent the background level measured with the
  empty target and normalized on the \isotope[10]{Be} beam integral.
  The arrows mark \isotope[9]{Li} first excited state with \(J^\pi=1/2^-\) and
  \(E^*=2.69\)~MeV.
\item The differential cross-sections for the
  \(\isotope[2]{H}(\isotope[10]{Be},\isotope[3]{He})\isotope[9]{Li}\) reaction.
  Points show
  the angular distribution  for the
  \(\isotope[2]{H}(\isotope[10]{Be},\isotope[3]{He})\isotope[9]{Li}_\text{g.s.}\)
  reaction obtained in the experiment. 
  (The error bars are statistical only.)
  The solid curve shows the result of corresponding
  FRESCO one-step calculations, see text for the details.
\item The inclusive MM spectrum of \isotope[8]{Li} from the
  \(\isotope[2]{H}(\isotope[10]{Be},\isotope[4]{He})\isotope[8]{Li}\)
  reaction.
  The gray filled histogram corresponds to the background measurement with
  the empty target normalized on the beam integral.
\item (\textit{a}) The excitation spectra of \isotope[8]{Li} obtained in the case
  of the \isotope[4]{He} recoil in coincidence with \isotope[8]{Li}
  (filled histogram) and \isotope[7]{Li} (black histogram).
  The arrows mark the excitation energy for the ground (\(2^+\)), first (\(1^+_1\)),
  second (\(3^+\)), third (\(1^+_2\)) \isotope[8]{Li} excited states, and
  the \isotope[8]{Li} neutron separation energy (\(S_n\)).
  (\textit{b}) The excitation spectrum of \isotope[8]{Li} obtained from the
  \(\isotope[2]{H}(\isotope[10]{Be},\isotope[4]{He})\isotope[8]{Li}\)
  reaction data after the empty-target subtraction with the condition of
  coincidences \isotope[4]{He}--\isotope[7]{Li} (black histogram)
  and \isotope[4]{He}--\isotope[8]{Li} (filled histogram).
  The solid curve corresponds to the Monte-Carlo calculation for the
  contribution of the 2.255 MeV (\(3^+\)) \isotope[8]{Li} state.
\item The FRESCO calculation of the differential cross sections for the
    \(\isotope[2]{H}(\isotope[10]{Be},\isotope[4]{He})\isotope[8]{Li}\)
    reaction populating the \(3^+\) state with \(E^*=2.255\)~MeV (solid line) and
    the \(1^+_2\) state  with \(E^*=3.21\) MeV (dashed line) of \isotope[8]{Li}
    (only pure main components of \(\isotope[10]{Be}_{\text{g.s.}}\) and
    \isotope[8]{Li} wave functions were taken into account).
\end{enumerate}

\clearpage 

\end{document}